\documentclass[prl,aps,
twocolumn,
floatfix,
preprintnumbers,
superscriptaddress]{revtex4-1}

\usepackage{hyperref}
\usepackage[dvipsnames]{xcolor}
\usepackage{amsmath,amssymb,mathrsfs,amsthm,tikz,shuffle,paralist,accents}

\usetikzlibrary{shapes.misc}
\usetikzlibrary{cd}
\usetikzlibrary{snakes}
\usetikzlibrary{calc}
\usepackage{mathtools}
\usepackage{slashed}

\usepackage[all]{xy}

\tikzset{
    ncbar angle/.initial=90,
    ncbar/.style={
        to path=(\tikztostart)
        -- ($(\tikztostart)!#1!\pgfkeysvalueof{/tikz/ncbar angle}:(\tikztotarget)$)
        -- ($(\tikztotarget)!($(\tikztostart)!#1!\pgfkeysvalueof{/tikz/ncbar angle}:(\tikztotarget)$)!\pgfkeysvalueof{/tikz/ncbar angle}:(\tikztostart)$)
        -- (\tikztotarget)
    },
    ncbar/.default=0.5cm,
}
\tikzset{round left paren/.style={ncbar=0.5cm,out=115,in=-115}}
\tikzset{round right paren/.style={ncbar=0.5cm,out=60,in=-60}}

\newcommand{\npcb}{I^{\text{NPCB}}}
\newcommand{\diff}{\text{d}}
\newcommand{\lbl}{\text{LBL}}
\newcommand{\GZ}{\text{MOM}}
\newcommand{\sing}{{\sc Singular}}
\newcommand{\oz}{\hat{z}}
\newcommand{\oy}{\hat{y}}
\newcommand{\oa}{\hat{\alpha}}
\newcommand{\oP}{\hat{P}}

\newcommand{\vv}{v}
\newcommand{\vw}{w}

\DeclareMathOperator{\aut}{Aut}
\DeclareMathOperator{\raut}{\overline{Aut}}

\newcommand{\PGL}{\text{PGL}}
\newcommand{\hypel}{\mathcal{H}}

\newcommand{\bq}{\begin{eqnarray}}
\newcommand{\eq}{\end{eqnarray}}

\begin{document}

\preprint{CERN-TH-2023-133}
\preprint{MITP-23-033}
\preprint{ZU-TH 33/23}

\title{Genus Drop in Hyperelliptic Feynman Integrals}

\author{Robin~Marzucca}
\affiliation{Physik-Institut, Universit\"at Z\"urich, Winterthurerstrasse 190, 8057 Z\"urich, Switzerland}
\author{Andrew~J.~McLeod}
\affiliation{CERN, Theoretical Physics Department, 1211 Geneva 23, Switzerland}
\affiliation{Mani L. Bhaumik Institute for Theoretical Physics, Department of Physics and Astronomy,
UCLA, Los Angeles, CA 90095, USA}
\author{Ben~Page}
\affiliation{CERN, Theoretical Physics Department, 1211 Geneva 23, Switzerland}
\author{Sebastian~P\"ogel}
\affiliation{PRISMA Cluster of Excellence, Institut f\"ur Physik, Staudinger Weg 7, \\
Johannes Gutenberg-Universit\"at Mainz, D - 55099 Mainz, Germany}
\author{Stefan Weinzierl}
\affiliation{PRISMA Cluster of Excellence, Institut f\"ur Physik, Staudinger Weg 7, \\
Johannes Gutenberg-Universit\"at Mainz, D - 55099 Mainz, Germany}

\begin{abstract}
The maximal cut of the nonplanar crossed box diagram with all massive internal propagators was long ago shown to encode a hyperelliptic curve of genus 3 in momentum space. 
Surprisingly, in Baikov representation, the maximal cut of this diagram only gives rise to a hyperelliptic curve of genus 2.
To show that these two representations are in agreement, we identify a hidden involution symmetry that is satisfied by the genus 3 curve, which allows it to be algebraically mapped to the curve of genus 2. 
We then argue that this is just the first example of a general mechanism by means of which hyperelliptic curves in Feynman integrals can drop from genus $g$ to $\lceil g/2 \rceil$ or $\lfloor g/2 \rfloor$, which can be checked for algorithmically. 
We use this algorithm to find further instances of genus drop in Feynman integrals.
\end{abstract}

\maketitle

%************************************************************************************************************************************************%

\vspace{.14cm}
\noindent {\bf Introduction}
\vspace{.1cm}

Our ability to compute scattering amplitudes has advanced tremendously in recent years. Much of this progress has stemmed from our theoretical control over multiple polylogarithms~\cite{Chen,G91b,Goncharov:1998kja,Remiddi:1999ew,Borwein:1999js,Moch:2001zr,FBThesis,Goncharov:2010jf,2011arXiv1101.4497D,Panzer:2014caa}, which turn out to be sufficient for expressing all amplitudes at one loop and certain classes of amplitudes to all orders~\cite{Caron-Huot:2011zgw,Dixon:2016nkn,Bourjaily:2018aeq,Caron-Huot:2018dsv,Drummond:2018caf,Caron-Huot:2019vjl,He:2020vob,He:2021non,Dixon:2022rse}. Beyond one loop, however, it remains unclear what classes of functions can appear in amplitudes, even at fixed loop order. Our continued progress thus crucially depends on begin able to characterize the types of functions that appear, and in particular on our ability to reliably diagnose the minimal class of functions a given Feynman integral can be evaluated in terms of.

One of the types of functions that are known to arise starting at two loops are integrals over hyperelliptic curves~\cite{Huang:2013kh,Georgoudis:2015hca,Doran:2023yzu}. 
Hyperelliptic curves are algebraic curves of genus $g>1$ whose defining equation takes the form $y^2=P(z)$, for some polynomial $P(z)$ of degree $(2g+1)$ or $(2g+2)$.
They generalize elliptic curves, whose defining equation takes the same form when $g=1$. Feynman diagrams that give rise to elliptic curves have received significant attention in recent years~\cite{SABRY1962401,Broadhurst:1993mw,Caffo:1998du,Laporta:2004rb,Muller-Stach:2011qkg,Paulos:2012nu,Caron-Huot:2012awx,Nandan:2013ip,Adams:2013kgc,Adams:2013nia,Bloch:2013tra,Remiddi:2013joa,Adams:2014vja,Adams:2015gva,Adams:2015ydq,Bloch:2016izu,Remiddi:2016gno,Adams:2016xah,Bonciani:2016qxi,vonManteuffel:2017hms,Adams:2017ejb,Bogner:2017vim,Ablinger:2017bjx,Chicherin:2017bxc,Bourjaily:2017bsb,Broedel:2017siw,Adams:2018yfj,Broedel:2018iwv,Adams:2018bsn,Adams:2018kez,Broedel:2018qkq,Bourjaily:2018yfy,Mastrolia:2018uzb,Honemann:2018mrb,Broedel:2019hyg,Broedel:2019kmn,Abreu:2019fgk,Campert:2020yur,Frellesvig:2021vdl,Kristensson:2021ani,Morales:2022csr,McLeod:2023qdf,mirrors_and_sunsets},
as have diagrams that give rise to integrals over higher-dimensional varieties~\cite{Bloch:2014qca,Bloch:2016izu,Bourjaily:2018ycu,Bourjaily:2018yfy,Bourjaily:2019hmc,Klemm:2019dbm,Vergu:2020uur,Bonisch:2020qmm,Bonisch:2021yfw,Pogel:2022yat,Duhr:2022pch,Pogel:2022ken,Pogel:2022vat,Duhr:2022dxb,Cao:2023tpx,McLeod:2023doa}. 
In contrast, Feynman diagrams that give rise to curves with genus $g>1$ have received little attention. A more in-depth study of these integrals is long overdue.

\begin{figure}[t]
\centering
\includegraphics{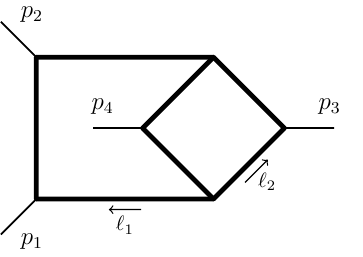}
\caption{The nonplanar crossed box diagram, with massive internal propagators.}
\label{fig:nonplanar_crossed_box}
\end{figure}

The types of integrals that appear in Feynman diagrams can be diagnosed by studying their maximal cut, in which all propagators are put on-shell.
In this letter, we focus on diagrams whose maximal cut is given by a one-fold integral of the form
\bq
 \int \frac{dz \, N(z)}{\sqrt{P\left(z\right)}} \, ,
\eq
where $N(z)$ and $P(z)$ are polynomials of $z$~\footnote{More specifically, we consider holomorphic integrands,
which implies that the degree of $N(z)$ should satisfy $3+2 \deg N \le \deg P$.}, 
as these diagrams are expected to give rise to iterated integrals involving differential one-forms related to the curve $y^2=P(z)$~\footnote{Note that Feynman integrals will also in general depend on the integral geometries associated with the maximal cut of their subtopologies.}.
In particular, we highlight that the curve one associates with a given Feynman diagram in this way can have different genus depending on the integral representation one studies.
We illustrate this with the example of the massive nonplanar crossed box diagram shown in Figure~\ref{fig:nonplanar_crossed_box}; while the maximal cut of this diagram involves a curve of genus 3 in momentum space~\cite{Georgoudis:2015hca}, its maximal cut in Baikov representation only involves a curve of genus 2. 

In order to make sense of this discrepancy, we analyze the properties of these
curves in more detail. We find that the genus 3 curve has an extra involution
symmetry, which can be made manifest through a change of variables that brings
its defining equation into a form that only depends on $z^2$ (whereupon the
action of the involution is simply $z \rightarrow -z$). Through the change of
variables $w = z^2$, this genus 3 curve can then be mapped to a curve of genus 2 that is isomorphic to the curve that appears in the Baikov representation. 

More generally, we show that any genus $g$ hyperelliptic curve that enjoys an extra involution symmetry can be related to a pair of hyperelliptic curves of genus $\lceil \frac{g}{2} \rceil$ and $\lfloor \frac{g}{2} \rfloor$. After constructing the explicit mappings between these curves, we present an algorithm for detecting the existence of such an involution and constructing the change of variables that makes this involution manifest. Finally, we describe the implications that this extra involution has for the period matrices of these hyperelliptic curves, and show that the entries of this matrix satisfy linear relations among themselves. We conclude by pointing to additional examples of Feynman integrals whose maximal cuts in momentum space experience similar genus drops.

%************************************************************************************************************************************************%

\vspace{.14cm}
\noindent {\bf The Nonplanar Crossed Box}
\vspace{.1cm}

The main example we focus on in this letter is the nonplanar crossed box diagram in four dimensions, shown in Figure~\ref{fig:nonplanar_crossed_box}.
It can be written as
\begin{equation}
\label{eq:example_Feynman_integral}
	\npcb = \int\frac{\diff^4\ell_1}{i\pi^2}\frac{\diff^4\ell_2}{i\pi^2}\frac{1}{\prod_{i=1}^{7}D_i} \, ,
\end{equation}
where the propagators are given by
\begin{equation}
	\begin{gathered}
		D_1=\ell_1^2-m_1^2 \,,\quad D_2=(\ell_1-p_1)^2-m_2^2\, , \\
		D_3=(\ell_1-p_1-p_2)^2-m_3^2\,,\quad D_4=\ell_2^2-m_4^2 \, ,\\
		D_5=(\ell_2-p_3)^2-m_5^2\,,\quad D_6=(\ell_1+\ell_2)^2-m_6^2 \,,\\
		D_7=(\ell_1+\ell_2-p_1-p_2-p_3)^2-m_7^2 \, .
	\end{gathered}
\end{equation}
The external momenta are massless with $p_i^2=0$.
This integral depends on between three and nine kinematic variables, depending on how many of the internal masses are taken to be distinct. 
For the majority of our analysis, it will be sufficient to take all masses to be identical (but nonzero), $m_i = m$. 
In that case, the integral depends on just $s = (p_1+p_2)^2$, $t= (p_2+p_3)^2$, and $m^2$. 

In~\cite{Huang:2013kh,Georgoudis:2015hca}, the maximum cut of this diagram was computed directly in momentum space, and was found to take the form
\begin{equation} \label{eq:momentum_max_cut}
    \text{MaxCut}_{\GZ}(\npcb) \sim \int \frac{\diff z\ z}{\sqrt{P_{8}(z)}} \, ,
\end{equation}
where $P_{8}(z)$ is a polynomial of degree eight in the variable $z=\text{tr}_-(p_4 p_2 \ell_1 p_1)/s^2$ whose coefficients depend on $s$, $t$, and $m_i^2$~\footnote{We define \unexpanded{$\text{tr}_\pm(abcd)\equiv\text{tr}(\frac{1\pm\gamma_5}{2} \slashed{a}\slashed{b}\slashed{c}\slashed{d})$}}.
The maximal cut thus defines a period of a hyperelliptic curve of genus 3~\footnote{This genus can also be confirmed by feeding the seven on-shell equations directly into \sing~\cite{DGPS}.}.
This implies the full integral should be expressible in terms of iterated integrals involving one-forms related to the genus 3 curve defined by $P_8(z)$.
 
We can also compute the maximal cut of this integral using a loop-by-loop Baikov parametrization. 
In this representation, the maximal cut takes the form
\begin{equation}
    \text{MaxCut}_{\lbl}(\npcb) \sim \int \frac{\diff z}{\sqrt{P_{6}(z)}} \, ,
\end{equation}
where now the integration variable $z=(\ell_1\cdot p_3)$ is an irreducible scalar product that was introduced when realizing the Baikov representation, and $P_{6}(z)$ is a degree-six polynomial in $z$. 
As a result, this form of the maximal cut seems to imply that the nonplanar crossed box can be expressed in terms of iterated integrals involving one-forms related to a curve of genus 2. 
We have also computed the Picard--Fuchs operator of the maximal cut on various generic one-dimensional kinematic slices, and find that these operators are always of order 4, consistent with a curve of genus 2.

This leaves us in a perplexing situation, as our expectation for the types of iterated integrals that will appear in the nonplanar crossed box depends on which representation of the integral we start from. 
Already in the equal-mass case, where 
\begin{widetext}
\begin{equation} \label{eq:curves}
\begin{aligned}
	P_{6}(z)&=
		s \left(2 z (s+2 z)-3 m^2 s\right) \left(m^2s+2 z (s+2 z)\right) \left(s (s+t+2 z)^2-4 m^2t (s+t)\right) \, ,\\[.1cm]
	P_{8}(z)&=(s+t)^2 \left(t^2 m^2+s^2 z (s z+t)\right) \left(m^2 (s+t)^2+s^2 z (s z+s+t)\right)\times \\
	&\hphantom{{}=}\left(s^2 z m^2 \left(-3 s^3 z+s^2 (2 t z+t)+s t^2 (2 z+3)+2 t^3\right)+t^2 \left(m^2\right)^2 (s+t)^2+s^4 z^2 (s z+t) (s z+s+t)\right) \,,
    \end{aligned}
\end{equation}
\end{widetext}
we find curves of different genus, between which there can exist no birational transformation.

%************************************************************************************************************************************************%

\vspace{.14cm}
\noindent {\bf A Mechanism for Genus Drop}
\vspace{.1cm}

The resolution of this apparent discrepancy turns out to be related to the group
of automorphisms that we can associate to the genus 3 curve. 
In particular, this curve has an additional involution symmetry that gives rise to relations between the entries of its period matrix. 
As a result, it can algebraically be mapped to a pair of curves of lower genus. 
To see how this comes about, we now review some properties of hyperelliptic curves. 

A hyperelliptic curve of genus $g$ is defined by the (affine) equation
\begin{equation}\label{eq:HyperEllipticGeneric}
	\hypel: y^2=P(z),
\end{equation}
where $P(z)$ is a polynomial of degree $2g+1$ or $2g+2$. We denote the
roots of $P(z)$ as $\alpha_i\in\mathbb{P}^1$, and without loss of generality we specalize to the
case where the degree is $2g+2$.
To study algebraic properties of $\hypel$, we consider the field of rational
functions on the curve, $\mathbb{C}(y,z)$: rational functions of $y$ and $z$
such that $y$ satisfies equation~\eqref{eq:HyperEllipticGeneric}.
Viewing $\mathbb{C}(y,z)$ as a field extension of $\mathbb{C}$ we 
can consider the group of automorphisms of this extension: 
$\aut{(\hypel)} \equiv \aut{(\mathbb{C}(y,z)\slash  \mathbb{C})}$. This is the
set of automorphisms of $\mathbb{C}(y,z)$ that act as the identity on
$\mathbb{C}$. 
This group $\aut{(\hypel)}$ naturally always includes the involution
\begin{equation}
	e_0:y\to -y \, ,
\end{equation}
while the reduced automorphism group 
$\raut{(\hypel)}\equiv\aut{(\hypel)}/\langle e_0\rangle$ is a subgroup of $\aut{(\mathbb{C}(z)/\mathbb{C})}\cong \text{PGL}_2(\mathbb{C}) $ and consists of 
M\"obius transformations that permute the roots of the polynomial $P(z)$~\cite{Gutierrez2005HyperellipticCO,Shaska_2014}.

Let us study the reduced automorphism group by considering coordinate redefinitions in $z$ via M\"obius transformations.
For $\smash{\gamma=\begin{psmallmatrix}a&b\\c&d\end{psmallmatrix}\in \PGL_2(\mathbb{C})}$, one finds a new representation of the curve 
\begin{equation}
	\hypel: \oy^2=\oP(\oz) \, ,
\end{equation}
where
\begin{equation}
	z=\gamma[\oz]\equiv\frac{a \oz +b }{c \oz +d} \, ,\quad y= \hat{y}\frac{1}{(c \oz+d)^{2g+2}} \, ,
\end{equation}
and 
\begin{equation}
	\oP(\oz)=(c\oz+d)^{2g+2}P(\gamma[\oz]) \, .
\end{equation}
The roots $\oa_i$ of $\oP$ are related to those of $P$ by
\begin{equation}
	\oa_i \equiv \gamma^{-1}[\alpha_i] = -\frac{d\alpha_i-b}{c\alpha_i-a} \, .
\end{equation}
A hyperelliptic curve is said to possess an extra involution
$e_1\in\raut{(\hypel)}$ if there exists a $\gamma\in\PGL_2(\mathbb{C})$,
such that we have
\begin{equation}
\label{eq:roots_in_pairs}
\oP(\oz) = Q(\oz^2) \equiv c(\oz^2-\oa_1^2)\ldots(\oz^2-\oa_{g+1}^2) \, ,
\end{equation}
where $c\in\mathbb{C}$ is a constant~%
\footnote{If a $\PGL_2(\mathbb{C})$ transformation exists to bring a curve into the form of equation~\eqref{eq:roots_in_pairs}, it will not be unique; alternate transformations exist that lead to the same projective curve in different affine charts.}.
The roots thus come in pairs $\pm\oa_i$, and the extra involution acts as
\begin{equation}
	e_1: \oz\to -\oz \, .
\end{equation}
We can then define another involution by composition,
\begin{equation}
e_2=e_1\circ e_0:(\hat y,\hat z)\to (-\hat y,-\hat z).
\end{equation}
To our curve $\mathcal{H}$, we can associate the two curves
\begin{equation}
	\begin{aligned}
		\hypel_1&:\vv_1^2= Q(\vw)=c(\vw-\oa_1^2)\ldots(\vw-\oa_{g+1}^2) \, ,\\
		\hypel_2&:\vv_2^2=\vw Q(\vw)=c\vw (\vw-\oa_1^2)\ldots(\vw-\oa_{g+1}^2) \, . \\
	\end{aligned}
\end{equation}
We can recover $\hypel$ via the maps
\begin{equation}
\begin{aligned}
	\rho_1&:(\vv_1,\vw)\to (\oy,\oz^2) \, ,\\
	\rho_2&:(\vv_2,\vw)\to (\oy\oz,\oz^2) \, ,
\end{aligned}
\end{equation}
which are invariant under $e_1$ and $e_2$ respectively.
From the degree of their defining polynomial in $\vw$, the curves $\hypel_1$ and $\hypel_2$
have genera $g_1=\lfloor\frac{g}{2}\rfloor$ and $g_2=\lfloor\frac{g+1}{2}\rfloor=\lceil\frac{g}{2}\rceil$.
Note that, in addition to the roots at $\oa_i^2$, the curve $\hypel_2$ has a branch point at 0; 
moreover, depending on whether $\lfloor\frac{g}{2}\rfloor$ is even or odd,
either $\hypel_1$ or $\hypel_2$ will have a branch point at $\infty$.

To check whether a given curve has such an extra involution, we can make a generic 
transformation $\gamma\in \PGL_2(\mathbb{C})$, and solve for the parameters $a,b,c,d$ such 
that the coefficients of odd powers of $\oz$ in $\oP$ vanish.
Alternatively, we can formulate a necessary and sufficient condition for the 
existence of such a transformation on the roots $\alpha_i$ of $P(z)$.
Consider the partitioning of the $(2g+2)$ roots into pairs 
$\smash{(p^{(i)}_1,p^{(i)}_2)}$ for $i=1,\ldots,g+1$. 
It can be shown that there exists a $\gamma\in\PGL_2(\mathbb{C})$ such that 
$\smash{\gamma^{-1}[p^{(i)}_1]=-\gamma^{-1}[p^{(i)}_2]}$ if and only if there exists a 
$\nu=(\nu_1,\nu_2,\nu_3)\in\mathbb{C}^3$ with $\nu_2^2 - 4\nu_1\nu_3\ne 0$ such that
\begin{equation}
	\begin{pmatrix}
		2p^{(1)}_1p^{(1)}_2 & p^{(1)}_1+p^{(1)}_2 & 2\\
		\vdots& \vdots &\vdots\\
		2p^{(g+1)}_1p^{(g+1)}_2 & p^{(g+1)}_1+p^{(g+1)}_2 & 2
	\end{pmatrix}\cdot\nu = \vec{0} \, , \label{eq:extra_involution_test}
\end{equation}
where the right-hand side corresponds to the $(g{+}1)$-dimensional zero vector. 
To check whether a curve has an extra involution we need to test 
this condition for all $(2g+1)!!$ pairings of the roots. 
If we find a vector $\nu$ that satisfies this condition, the transformations 
$\gamma$ that make the involution symmetry manifest are (up to an overall normalization) 
\begin{equation}
	\gamma=\left\lbrace
	\begin{array}{ll}
		\left(
		\begin{array}{cc}
		 \nu _2 & -\lambda\frac{\nu _3}{\nu _2} \\
 		0 & \lambda \\
		\end{array}
		\right), & \nu_1=0,\\
		\left(
\begin{array}{cc}
 \nu _2+\sqrt{\nu _2^2-4 \nu _1 \nu _3} & \lambda\frac{\sqrt{\nu _2^2-4 \nu _1 \nu _3}-\nu _2}{\nu _1} \\
 -2\nu_1 & 2\lambda \\
\end{array}
\right),& \nu_1\neq0 \, ,
	\end{array}
\right.
\end{equation}
where $\lambda\in\mathbb{C}\setminus\{0\}$ is the projective degree of freedom
of~\cite{Note5}.

%************************************************************************************************************************************************%

\vspace{.14cm}
\noindent {\bf Genus Drop in the Maximal Cut}
\vspace{.1cm}

We now illustrate how this genus drop mechanism works for the case of the non-planar crossed box. Starting from a general $\PGL_2(\mathbb{C})$ transformation $z=\gamma_1[\oz]$ and solving for the parameters of this transformation such that the coefficients 
of odd power monomials in $\oz$ vanish, we find a transformation
\begin{equation}
\label{eq:genus_3_symmetry_transform}
	\gamma_1=\left(\begin{array}{cc}
 		1& -1\\
 		r&r
	\end{array}\right),\quad 
	z=\gamma_1[\oz]=\frac{1}{r}\frac{\oz-1}{\oz+1},
\end{equation}
where $r=\sqrt{s^3/(m^2 t(s+t))}$. 
The polynomial
\begin{equation}
	Q_4(\oz^2)=(r\oz+r)^8\ P_8\left(\frac{1}{r}\frac{\oz-1}{\oz+1}\right)
\end{equation}
then only depends on powers of $\oz^2$. (Note that $Q_4(\oz^2)$ is of degree 4 in $\oz^2$, and therefore of degree 8 in $\oz$.) 
In this case, the extra involution maps
\begin{equation}
e_1\big[\text{tr}_-(p_4 p_2 \ell_1 p_1)\big] = \text{tr}_+(p_4 p_2 \ell_1 p_1) \, .
\end{equation} 
Hence, the extra involution can be associated to parity.

Following the notation introduced in the previous section, we define $w\equiv\oz^2$, $v_1\equiv y$, and $v_2 \equiv y\,w$.
We can then associate two curves with the original genus 3 curve:
the elliptic curve
\begin{equation}
	v_1^2 = Q_4(w)
\end{equation}
and the genus 2 hyperelliptic curve 
\begin{equation}
	v_2^2 = w Q_4(w)\equiv Q_5(w).
\end{equation}
By considering the period matrices of the latter curve, one can show that its
period matrix generates the same lattice as the curve found from the Baikov
representation, and is therefore isomorphic.
Equivalently, we find that these curves have the same absolute invariants (as defined by Clebsch or Igusa~\cite{Igusa_1960,Clebsch_1967,Beshaj_2018}), which
characterize curves of genus 2
\footnote{
For curves of genus 3 defined over a field of characteristic 0, similar invariants were found by Shioda~\cite{Shioda_1967,Shaska_2014}. 
We include code for calculating these genus 2 and genus 3 absolute invariants in an ancillary file.
}.
We can therefore find the explicit transformation,
\begin{equation}
	\gamma_2 = \begin{pmatrix}a&b\\c&d\end{pmatrix} = \tfrac{m^4}{\sqrt[3]{2s^5(s+t)} }\left(
\begin{array}{cc}
	1 & \frac{s+t}{2m^2} - \sqrt{\frac{(s+t)t}{sm^2}}\\
	1 & \frac{s+t}{2m^2} + \sqrt{\frac{(s+t)t}{sm^2}}
\end{array}
\right),
\end{equation}
which allows us to relate these curves via
\begin{equation}
	(c w+d)^6 Q_5(\gamma_2[w])=P_6(w)\,.
\end{equation}

%************************************************************************************************************************************************%

\vspace{.14cm}
\noindent {\bf Additional Period Relations}
\vspace{.1cm}

\begin{figure}
\centering
	\includegraphics[scale=0.75]{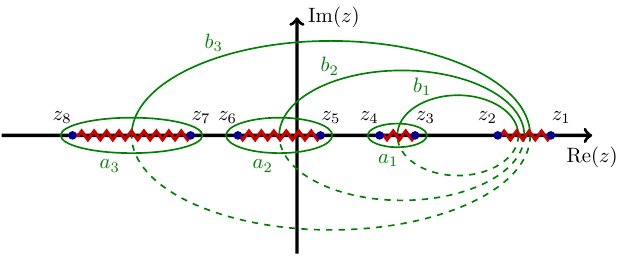}
	\includegraphics[scale=0.75]{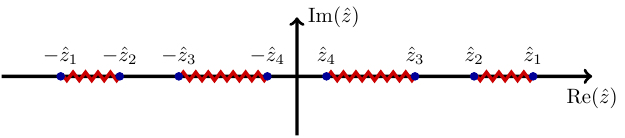}
\caption{The upper figure shows the branch cuts of a $g=3$ hyperelliptic curve and a symplectic basis of homology cycles.
The lower figure illustrates how roots pair up when this curve can be put in the form of equation~\eqref{eq:roots_in_pairs}.}
\label{fig:branch_cuts}
\end{figure}
The presence of the extra involution also has consequences for the periods of a hyperelliptic curve.
We recall that the periods of a curve correspond to the different possible pairings between a basis of holomorphic differentials $\omega_i$ and contours $\Gamma_j$.
For a hyperelliptic curve of genus $g$ defined by the polynomial $P_{2g + 2}(z)$, a basis of holomorphic differentials is given by $z^{i}\diff z/\sqrt{P_{2g+2}(z)}$ for $i=0,\ldots,g-1$.
Furthermore, one can find a symplectic basis of the homology, which is a basis of $2g$ contours $a_j$ and $b_j$ for $j=0,\ldots,g$, with the property that their intersection product is $a_i\circ b_j=\delta_{ij}$ and $a_i\circ a_j=b_i\circ b_j=0$.
An example of such a basis for a genus 3 curve with real roots is shown in Figure~\ref{fig:branch_cuts}.
The entries of the $g\times 2g$ dimensional period matrix $\mathcal{P}$ of the curve $\hypel$ are given by
\begin{equation}
	\mathcal{P}_{ij} = \int\limits_{\Gamma_j} \frac{z^{i}\diff z}{\sqrt{P_{2g+2}(z)}},
 \hspace*{2mm}
 \Gamma_j \in (a_1,\dots,a_g,b_1,\dots,b_g).
\end{equation}
Let us assume that $\hypel$ has an extra involution, which can be made
manifest using the coordinate transformation $z=\frac{a \oz+b}{c \oz +d}$. Then, 
\begin{align}
	&\int\frac{\diff z\ z^i}{\sqrt{P_{2g+2}(z)}} \nonumber\\
	=&\pm\int\frac{\diff w (ad-bc) (\pm a\sqrt{w}+b)^i (\pm c\sqrt{w} +d)^{-i+(g-1)}}{2\sqrt{w Q_{g+1}({w})}}, \label{eq:period_relation}
\end{align}
where 
$\hat{z}=\pm \sqrt{w}$.
Expanding out equation~\eqref{eq:period_relation}, we see that all of the periods of $\hypel$ are expressible as linear combinations of the periods of
subcurves $\hypel_1$ and $\hypel_2$~\footnote{In the example of the nonplanar
  crossed box, the maximal cut from equation~\eqref{eq:momentum_max_cut} gets
  mapped by~\eqref{eq:genus_3_symmetry_transform} to a linear combination of
  only genus 2 periods, since all non-integer powers of $w$ drop out of the
  numerator.}. More generally, we can find matrices $M_\omega\in
\mathbb{C}^{g\times g}$ and $M_\Gamma \in \mathbb{Z}^{2g\times 2g}$ such that
\begin{equation}
	\label{eq:period-matrix-split}
	M_\omega^t \mathcal{P} M_\Gamma=
	\begin{pmatrix}
		\mathcal{P}_{1} & 0\\
		0 & \mathcal{P}_{2}
	\end{pmatrix} \, ,
\end{equation}
where $\mathcal{P}_1$ and $\mathcal{P}_2$ are period matrices of curves isogenous to $\hypel_1$ and $\hypel_2$.
This reflects the fact that the Jacobian variety $\text{Jac}(\hypel)$, which corresponds to the $2g$ dimensional lattice spanned by the column vectors of $\mathcal{P}$, is in this case isogenous to $\text{Jac}(\hypel_1)\times\text{Jac}(\hypel_2)$~\cite{Gutierrez2005HyperellipticCO,Cardona}.

Equation~\eqref{eq:period_relation} can also be used to find relations between the entries of $\mathcal{P}$.
For example, consider the genus 3 curve defined by $P_8(z)$ from equation~\eqref{eq:curves}:
the transformation $\gamma_1$ from equation~\eqref{eq:genus_3_symmetry_transform} can be inserted into~\eqref{eq:period_relation} to find the relation 
\begin{equation}
\begin{aligned}
	\int_{a_1} \frac{\diff z}{\sqrt{P_{8}(z)}}
	=r^2
	\int_{a_2} \frac{\diff z\ z^2}{\sqrt{P_{8}(z)}} \, .
\end{aligned}
\end{equation}
As can be deduced from the block-diagonal form of equation~\eqref{eq:period-matrix-split}, we expect there to be $4 \times \lfloor \tfrac{g}{2}\rfloor \times\lceil \tfrac{g}{2}\rceil$ relations between the periods of a curve with an extra involution.
%************************************************************************************************************************************************%

\vspace{.14cm}
\noindent {\bf Further examples}
\vspace{.1cm}

\begin{figure}[t]
\centering
\includegraphics{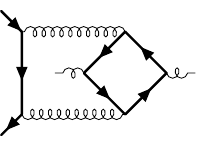}
\includegraphics{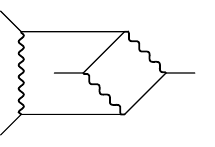}
\caption{Examples of hyperelliptic Feynman integrals in which genus drop via an extra involution can be observed. These integrals contribute to $gg\to t \overline{t}$ with a top loop, and M{\o}ller scattering $e^-e^-\to e^-e^-$ with the exchange of three $Z$ bosons.}
\label{fig:nonplanar_crossed_box_further_examples}
\end{figure}

Although we have mainly focused in this letter on the nonplanar crossed box with equal internal masses, genus drop can be observed in phenomenologically-relevant examples involving different masses. 
For example, Figure~\ref{fig:nonplanar_crossed_box_further_examples} depicts one
diagram that contributes to $t \overline{t}$ production and another that
contributes to M{\o}ller scattering. 
The maximum cut of both diagrams involve curves of genus 3 in momentum space that enjoy an extra involution symmetry and can be mapped to curves of genus 2. 

It is also possible to see genus drop via the same mechanism in special kinematic limits. 
For instance, a further involution symmetry appears in the equal-mass nonplanar crossed box diagram when $s=-2t$. 
In this limit there is a permutation symmetry that exchanges $p_1 \leftrightarrow p_2$, and the curve becomes 
\bq
\label{eq:curve_on_kinematic_slice}
 y^2
 = 
 8 t^2 
 \left( 2 \hat{z}^2 \!-\! m^2 t \right)
 \left( 4 \hat{z}^2 \!-\! 2 m^2 t \!-\! t^2 \right)
 \left( 4 \hat{z}^2 \!+\! 6 m^2 t \!-\! t^2 \right),
 \nonumber
\eq
with $\hat{z}=z-\frac{t}{2}$. This makes it evident that the maximal cut of this diagram drops from genus 2 to genus 1 in this limit~\footnote{We have also checked that the momentum 
and Baikov representations still define genus $3$ and genus $2$ curves when $s=-2t$, so this genus drop is not automatic.}.  
This is consistent with the Picard--Fuchs operator associated with this integral, which we also observe to drop from order 4 to order 2 when $s=-2t$.

\begin{figure}[t]
\centering
\includegraphics{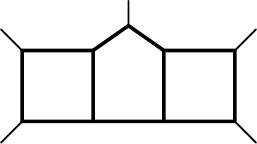}
\caption{The three-loop box-pentagon-box integral with equal internal masses and massless external momenta, which exhibits a genus drop from 5 to 3.}
\label{fig:box-penta-box}
\end{figure}

Finally, a similar genus drop can be observed for the five-point box-pentagon-box integral shown in Figure~\ref{fig:box-penta-box}, for massless external particles and equal internal masses. 
In momentum space, we find that the maximal cut of this integral gives rise to a curve of genus 5 using \sing~\cite{DGPS}, which matches our expectations from the results of~\cite{Huang:2013kh}. 
We also find that the maximal cut obtained using a loop-by-loop Baikov parametrization can be identified with a period of a hyperelliptic curve of genus 3. 
Notably---unlike the other examples we have considered---the momentum space curve is in this case not hyperelliptic; even so, we expect that a mechanism similar to the one we have described for hyperelliptic curves is responsible for this genus drop.

%************************************************************************************************************************************************%

\vspace{.14cm}
\noindent {\bf Conclusion}
\vspace{.1cm}

In this letter we have studied Feynman diagrams that give rise to integrals over
hyperelliptic curves, and highlighted the fact that different integral
representations of these diagrams can lead to curves with different genera. 
Importantly, this drop in genus represents a significant simplification in the
types of functions that these diagrams are expected to evaluate to.
In all of our hyperelliptic examples, we have observed that discrepancy in genus can be 
explained by the presence of an extra involution symmetry that allows the higher-genus curve to be algebraically mapped to the curve with lower genus.
We expect that the presence of extra involutions in the momentum representation can follow from
discrete Lorentz symmetries (spacetime parity or time reversal). We also presented an algorithm to detect when an extra involution exists, and showed that this symmetry leads to 
linear relations among the periods of the corresponding curve. 

While it is important to be able to diagnose which class of special functions a
given Feynman integral is expected to be expressible in terms of, it will be even more essential to develop the technology for working with these classes of functions. 
Despite remarkable recent progress on iterated integrals 
involving elliptic curves (see~\cite{Bourjaily:2022bwx} for an overview), much less technology has currently been developed 
for iterated integrals over hyperelliptic curves (however, for recent work see~\cite{Enriquez:2021,Enriquez:2022,DHoker:2023vax}). 
The nonplanar crossed box with equal internal masses represents an ideal example on which to develop such technology, given that it only involves a curve of genus 2 and depends on two dimensionless variables~%
\footnote{Notably, the double box integral was also recently shown to be hyperelliptic for external kinematics in more than four dimensions~\cite{Doran:2023yzu}; 
however, the double box exhibits much more kinematic complexity, making it a less-than-ideal example.}. 

Having identified a novel class of simplifications that can occur in hyperelliptic Feynman integrals, it is natural to wonder whether analogous simplifications can occur in Feynman integrals 
that involve integrals over more general varieties, such as curves that are not hyperelliptic or higher-dimensional Calabi--Yaus. 
One way to search for evidence of such simplifications would be to look for unexpected relations between entries of the period matrix. 
We leave this enticing possibility to future work.

%************************************************************************************************************************************************%

\vspace{.14cm}
\noindent {\bf Acknowledgments}
\vspace{.1cm}

We thank Alessandro Georgoudis, Oliver Schlotterer, and Yang Zhang for stimulating discussions and comments on the manuscript. 
This work has been supported by the European Union’s Horizon 2020 research and innovation program \textit{EWMassHiggs} (Marie Skłodowska Curie Grant agreement ID: 101027658).
SP and SW have been supported by the Cluster of Excellence Precision Physics, Fundamental Interactions, and Structure of
Matter (PRISMA EXC 2118/1) funded by the German Research Foundation (DFG) within
the German Excellence Strategy (Project ID 39083149).
RM, SP, and SW are grateful for the kind hospitality of the CERN Theory Department where part of this work was carried out.

%************************************************************************************************************************************************%

\bibliography{genus_drop}

\end{document}